\documentclass[%
 reprint,
 amsmath,amssymb,
 aps,
]{revtex4-2}

\usepackage{graphicx}
\usepackage{dcolumn}
\usepackage{bm}

\usepackage{parskip}

\newcommand{\quotepar}[2]{\hangindent=0.7cm \textbf{#1}: \textit{#2}}

\begin{document}

\preprint{APS/123-QED}

\title{Student Representations of Computation in the Physics Community}

\author{W. Brian Lane}
 \email{Brian.Lane@unf.edu}
\author{Cortney Headley}%
\affiliation{%
 Department of Physics, University of North Florida\\
 1 UNF Drive, Jacksonville, FL, 32224
}%

\date{\today}

\begin{abstract}

Undergraduate physics education has greatly benefited from the introduction of computational activities. However, despite the benefits computation has delivered, we still lack a complete understanding of the computationally integrated learning experience from the student perspective. In particular, we are interested in investigating how students develop expert-like perceptions of computation as a practice within the professional physics community. To investigate this aspect of student development, we employ the Communities of Practice framework, which describes how students navigate through a professional community by appropriating the community's practices and goals as personally important. We introduce the construct of a COP-Model as a student's internal representation of a professional community that they develop as they navigate that community. We used this construct to formulate a set of research questions and semistructured interview protocols to explore how five physics students represent the use of computation in their mental models of the global physics community. We foreground these representations in the local academic community established by their instructors in three concurrent computationally integrated physics courses. We find that these students saw computation as a normal and valuable part of physics practice, identified benefits of using computation in alignment with the physics community, struggle with confidence with regards to computation, and demonstrate some expectations for computational proficiency that are misaligned with the physics community. We establish these themes with interview excerpts and discuss implications for instruction and future research.

\end{abstract}

\maketitle


\section{\label{sec:IntroComp}Computation in Physics Education}



The use of computation in undergraduate physics education is on the rise for various reasons. Students can use computation to simulate physical systems, conduct advanced analysis of experimental data, create insightful visualizations, explore analytically intractable problems, and bridge the gap between mathematical problems and experimental activities \cite{weller2018investigating,Serbanescu11Putting,chabay08computational,young2019identifying,weber2021conceptual,singh2008interactive,mckagan2008developing,zwickl17characterizing,wang2021integrating,chonacky2008integrating}. Upon graduating, students find that computation is prevalent in all fields of physics research and STEM industry \cite{wolfram2002new,chonacky2008integrating,Burke17Developing,fracchiolla2021computational}, with faculty mentors and employers expecting well-developed computational skills in their research mentees and new hires \cite{thornton2009computational,caballero15skills,enrique2018computational,Graves20Hitting,zwickl17characterizing}. Integrating computation into the undergraduate experience provides students with research- and industry-relevant skills \cite{martin2016undergraduate,Graves20Hitting,fracchiolla2021computational} while helping them develop an appropriate view of building models and implementing them with a computer \cite{Aiken12Understanding}. Computation enables students to pursue creative solutions \cite{bing2008symbolic,Caballero14Model}, more directly engage in sense-making \cite{wieman2008oersted,singh2008interactive,sand18Computation,weber2021conceptual}, and test a variety of model-based predictions \cite{Buffler08Model,Caballero14Model,obsniuk15case,wang2021integrating}. Such activities equalize student mathematical backgrounds \cite{feurzeig2011programming,POPAT2019365,Rodriguez20CT}, promote deeper learning \cite{Caballero14Model,bott2020student,Dwyer13Computational,fracchiolla2021computational}, and support underrepresented groups \cite{orton2016bringing,stump2011collaborative,van2020equity,wang2021integrating}. In physics and beyond, computational thinking has become its own suite of learning objectives, including data practices, modeling and simulation practices, computational problem solving practices, and systems thinking \cite{chonacky2008integrating,weintrop2016defining,Dwyer13Computational,bott2020student,orban2020computational,weller2021developing,fracchiolla2021computational,wang2021integrating}.

The physics education community is steadily developing a wealth of resources for integrating computation into the curriculum at all levels. Banks of computational activities are available \cite{lane21analysis} to immediately integrate into courses of all topics and contexts. A common practice is to provide students with a minimally working program \cite{oleynik2019scientific,lunk2012framework,weatherford12student,weatherford11student} to jump-start the computational activity and avoid losing student engagement when developing code from scratch. Pedagogical strategies such as pair programming \cite{lye2014review,Zhong17Pair} promote a unique form of collaboration \cite{weller2021developing} as students use code to study models together. Assessing computationally integrated learning can take many forms, such as computationally integrated homework problems and term projects \cite{Caballero14Model,Burke17Developing,wang2021integrating}, computational essays \cite{odden19computational,odden2019physics}, integrated code-notebook systems \cite{cardoso2018using,blank2019nbgrader,yaniv2018simpleitk}, or integrated lab reports with experiment-based work \cite{chabay08computational,Serbanescu11Putting}. Observational studies are being conducted to identify what new challenges computation introduces to students \cite{rowatt17investigating} and how students tend to move through computational activities \cite{weller2021developing} so that the benefits of computation can be realized without additional interference.


The progressive development of computationally integrated teaching has prompted many questions that remain open: For example, how do we know whether computational integration is accomplishing all the promises outlined above? What pedagogical approaches promote student engagement with the computational model, and what approaches inadvertently provoke student apprehension towards programming \cite{lunk2016attitudes}? What aspects of programming do students need to understand in order to learn by working with a computational model? How can we properly scaffold a diverse set of programming skills alongside the preexisting objectives of the physics curriculum, and how do we make room for computation in the already-packed curriculum  \cite{chonacky2008integrating,leary2018difficulties, Serbanescu11Putting,fracchiolla2021computational}? How can departments negotiate varying levels of faculty proficiency with computation? If computation provides a more direct link between model-building and sense-making, what role does mathematical manipulation play in the 21st-century physics classroom \cite{bing2008symbolic}? And (as we consider in this paper) how does our integration of computation develop students' perceptions of computation as a physics practice? In many ways, these questions are recapitulations of topics already being studied by physics education researchers in the absence of computation. As such, there are many resources we can draw on to study the impacts of computation on physics learning and determine best practices for design, delivery, and assessment of computational activities. 

In this work, we investigate the question, ``How do students develop expert-like perceptions of computation as a practice within the professional physics community?'' We foreground the students' perceptions in the community of practice within their coursework as brokered by their instructors. We are motivated to consider this question by conversations with students and faculty that seem to indicate that, although computation has proven invaluable \textit{to physics educators}, students do not always perceive the benefits that computation offers or the prominent place that computation holds in the physics community \cite{joglekar2020normalizing,gavrin2021preliminary,weller2021developing,caballero2012implementing}. The fact that students might not expect to encounter computation in their physics classes or see its usefulness compounds the trend of low motivation among physics students \cite{oliveira2013using,madsen2015physics,guido2018attitude}. Computationally integrated physics courses must attend to these motivational considerations \cite{lunk2016attitudes,leary2018difficulties}, as students tend to pursue learning activities more intently if the activity elicits their attention and appears relevant to their broader goals \cite{keller1987development,august11enhancing,ibanez2017empirical,gungor2010teaching}. 


In Section \ref{sec:PERBackground}, we review the Communities of Practice framework and how it can be used to study the effects of computation on the physics learning experience. In Section \ref{sec:COPModel}, we establish our conceptual framework for the current study, drawing on insights from the Communities of Practice framework \cite{irving2014conditions,quan2018interactions} 
and mental models \cite{greca2002mental}, arriving at the conclusion that a student's mental model of a community of practice plays an important role in the development of their interests in and trajectories within a field. In Section \ref{sec:ContextMethods}, we discuss the context of our study, present our research questions, and outline the interviews we conducted to study the role of computation in students' mental models of the physics community of practice. In Section \ref{sec:Interviews}, we present the insights obtained from these interviews and identify common themes. Finally, in Section \ref{sec:Discussion}, we answer our research questions, suggest implications for instruction, reflect on our use of mental models within Communities of Practice, and outline further research questions.

\section{\label{sec:PERBackground}Computation and the Communities of Practice Framework} 

In this section, we review how the Communities of Practice (COP) framework helps us understand how students develop as physicists. We are particularly interested in how the COP framework can be applied to the integration of computation, which is often framed within an explicit goal of preparing students for professional work in STEM fields. We review much of the PER work that has been conducted about computation, interpreting the results through a COP lens. Our goal in this section is to motivate our conceptual framework in Section \ref{sec:COPModel}, which forms the basis of our study.

\subsection{The Communities of Practice Framework}

The COP framework emphasizes how novices navigate their place within an existing professional culture by familiarizing themselves with that culture's common goals and conventional practices that fulfill those goals \cite{wenger1999communities,Wenger2000Communities,wenger2002cultivating}. These goals and practices are collectively referred to as the COP's \textit{sense of joint enterprise}. Based on \cite{ford2006chapter}, Quan, Turpen, and Elby define a practice in the scientific context as ``a set of activities that are embedded within and work toward the aims of a scientific community'' \cite{quan2018interactions}. They elaborate that the practices within a scientific COP must be meaningfully connected with each other and used with the purpose of helping to meet the community's goal. Irving and Sayre describe these practices as ``what counts'' as doing physics \cite{irving2014conditions}.

By adopting the joint enterprise of a COP, a novice member of the community begins structuring their \textit{lived identity} \cite{wenger1999communities,close2016becoming,PICUP-COP,quan2018interactions}, which the COP framework defines as ``a complex interplay between identity as a negotiated experience of self, a sense of membership, a learning trajectory, a nexus of multimembership, and a belonging defined globally but experienced locally'' 
\cite{irving2020communities}. This identity is connected with demonstrations of competence in the practices that the COP values \cite{irving2020communities} and embodies what the novice comes to believe that it means to be a practicing member of the COP \cite{Li13Identity}.

A novice's navigation of a COP is envisioned as moving along a trajectory \cite{quan2018interactions}. Wenger conceived of several possible trajectories: insider trajectories (remaining central), peripheral trajectories (accessing the community without becoming a full member), inbound trajectories (peripheral-to-central), boundary trajectories (between two or more COPs), and outbound trajectories (exiting the COP) \cite{wenger1999communities}. For example, in the context of physics education, an insider trajectory might be a physics professor leading a research group. A peripheral trajectory might be a pre-med student taking their required introductory courses in preparation for the MCAT. An inbound trajectory might be a junior physics major applying to physics graduate programs. A boundary trajectory might be an undergraduate double-majoring in physics and communication with the goal of becoming a science-focused journalist. An outbound trajectory might be an undergraduate changing majors from physics to mathematics. The population of this study (junior physics majors in upper-division undergraduate courses) leads us to focus this discussion on insider and inbound trajectories, but we recognize that the full suite of trajectories helps one understand a diversity of student experiences. 

Movement along an inbound trajectory ``is neither a linear nor smooth process'' \cite{quan2018interactions}, and is exhibited by the learner's adoption of the COP's goals and approaches, and their increasingly central and significant role within the community \cite{lave1991situating}. In moving along this trajectory, the learner's membership within the COP ``shap[es] their perceptions, values, and interactions with others'' \cite{close2016becoming}, and each step along an inbound trajectory is paved with the appropriation of a new practice \cite{irving2020communities,barab2000practice,irving2016identity}. A possible assessment of a learner's trajectory could be their appropriation or rejection of the community's practices and norms as embodied by central members \cite{irving2020communities,boylan2016deepening,boaler2000mathematics,nardi2003mathematics}.

We emphasize that, as each learner approaches a COP along an inbound trajectory, their trajectory is as unique as their background, culture, and personal identity. Making an analogy with spherical coordinates, while all inbound trajectories are characterized by $|\vec{r}| \rightarrow 0$ (with the center of the community at the origin), each trajectory can have a unique set of angular values $(\theta,\phi)$. If central members of the community tend to favor inbound trajectories that fall within a subset of angular values (for example, newcomers who enter the same way the existing central members did), their community's membership will not represent the makeup of the surrounding world.

Many learners will find themselves navigating multiple COPs concurrently, which Wenger called a \textit{nexus of multimembership} \cite{wenger1999communities,close2016becoming}. For example, a student pursuing a career teaching high school physics must navigate the COPs presented by their physics coursework, their education coursework, and the school in which they work as a pre-service teacher. It is similarly important to consider smaller COPs within a broader COP, such as distinguishing the condensed matter COP from the elementary particle COP within the broader physics community
. One can even bifurcate down to an individual class or research group as a COP which serves as a local picture of the global physics community \cite{irving2014conditions}. For the purposes of this paper, we define the COP of our study as the physics majors and instructors within a set of concurrent physics classes, which we view as a local representation of the global physics community \cite{irving2014conditions,irving2015becoming}. We assume that the insights gained from this scope might be applicable to physics subdomains or to other subjects.

As part of this navigation, the COP's sense of joint enterprise is constantly being renegotiated by its members, particularly as newer members move from peripheral positions to more central positions \cite{irving2020communities}. For example, a research group's focus or methods might change, or the set of problems that hold a subfield's interest might expand into new territory or migrate away from topics that have been sufficiently explored. We might say that the very existence of PER as a subfield of physics research is indicative of the expansion of the goals and approaches of the broader physics COP. The nexus of multimembership is particularly important for this renegotiation, so that new ideas can be applied to a community's existing purposes. For example, the introduction of computation to physics education is a renegotiation of the community's practices based on developments in the technology community.

\subsection{\label{subsec:COP-Physics}Communities of Practice in Physics Education}

From the perspective of the COP framework, an important overarching goal of the physics curriculum is for students to begin to align with the goals and approaches valued by the physics community through \textit{legitimate peripheral participation}: ``participation in practices in which a learner can engage that are socially warranted or legitimized by existing practitioners... and are appropriate for a newcomer'' \cite{irving2020communities}, leading the newcomer to pursue an inbound trajectory \cite{lave1991situated}. In this participation, students (ideally) develop their confidence in their ability to contribute to the community's goals using the community's practices. The student's learning is thus an adoption of shared understanding and practices \cite{irving2020communities,wenger2002cultivating,denscombe2008communities}. This participation takes place in the local community context (``What practices are important in this department, this course, or this research group?'') as an expression of the global physics community (``What practices are important in a given research field, as expressed in community artifacts like research papers and conference presentations?'').

This overarching goal is an example of one's lived identity as being defined globally (by the broader physics community) but experienced locally (within a particular course, curriculum, or research group). The local context, as a microcosm of the broader community, provides well-defined local engagement in the practices that are meaningful within the global COP \cite{irving2020communities}. 


The importance of the learning environment (in the classroom, a department, a curriculum) cannot be overstated in regards to helping students progress along their appropriate trajectories, particularly along an inbound trajectory that retains a student within the physics COP. For example, participation in undergraduate research (a key event in the physics learning experience) plays an important role in undergraduates' trajectories toward becoming practicing physicists \cite{irving2015becoming}. 
The availability of legitimate practices enables them to explore possible trajectories \cite{irving2016identity,lave1991situated}. For example, structural choices in a practice-oriented physics laboratory course can help establish the course as a local COP, motivate student engagement with that local COP, and establish students on inbound trajectories \cite{irving2014conditions}. Hamerski, Irving, and McPadden describe the development of a community of practice among Learning Assistants, who developed their own practices of feedback in partnership with faculty \cite{hamerski2021learning}. Similarly, physics students can develop their identity as a subject-matter expert through informal outreach to learners of varying backgrounds \cite{rethman2021impact}, providing a demonstration of their increasingly central membership in the physics community \cite{fracchiolla2020community}. Finally, the COP framework can be used to understand the experiences of underrepresented groups \cite{brickhouse2000kind,danielsson2012exploring,gonsalves2016masculinities}, as the lived identity developed in a COP is an important factor in student persistence in physics \cite{irving2014conditions}. Understanding and appreciating the breadth of possible student trajectories within our local and global physics communities is necessary for supporting students from underrepresented groups. 

\subsection{Understanding Computation through Communities of Practice}

The COP framework is particularly useful in studying the impact of computation on physics education because computation is so often taught as a set of recognized practices that physics researchers and STEM professionals use in pursuing a variety of goals and expect new graduates to be prepared to participate in \cite{Graves20Hitting,Burke17Developing}. Such practices include extracting computational insight, building computational models, debugging, data practices, demonstrating constructive dispositions toward computation, and working in groups on computational models \cite{weller2021developing}. These practices are established from both a theoretical perspective \cite{weintrop2016defining} and from pragmatic considerations of industry demands.

For example, Graves and Light \cite{Graves20Hitting} surveyed computational physicists to identify computational research skills they value as necessary in research students: converting a problem to a step-by-step procedure amenable to coding; flow control; proactive communication of results and issues; asking people for help with running or writing code; googling to answer coding questions; visualizing line data; and visualizing image data. They also explored how these professionals differentiate skills they expect students to develop in advance of beginning a research project from skills they expect students to develop over the course of their research. We interpret these different stages of skill development as being legitimate peripheral participation (in advance) and legtimeate participation closer to the center of the community (during research) \cite{irving2020communities}. They then compared these responses with the learning objectives of 4 exemplar computationally integrated undergraduate physics courses, highlighting similarities and differences between researchers' expectations (in the global professional community) and curricular opportunities (in the local academic community). 

Burke and Atherton \cite{Burke17Developing} interviewed STEM faculty and postdocs who actively use computation in their research, using emergent themes to establish a cycle of activities that computational research requires: Physical Transcription (gaining an initial understanding from analytical steps, choosing a mathematical model and computational implementation, articulating assumptions and goals), Planning (outlining code, identifying libraries to be imported and sections to be designed), Implementation (developing code with documentation), Testing (test each individual piece of the code), Running (using code to generate data with systematically varied inputs), Visualization (gaining intuition, deciphering results, and preparing communication of results), Numerical Analysis (identifying sources and impacts of errors, exploring stability and problem scaling), Physical Analysis (comparing results to hypotheses and verifying reasonability of results). They then used these themes to begin the backwards design of a computational physics course. Such a focused and results-oriented design process might alleviate frequent concerns that computation requires an unmanageable amount of space in the curriculum \cite{leary2018difficulties}.

Thornton, et al  \cite{thornton2009computational} and Enrique, Asta, and Thornton \cite{enrique2018computational} found among academic respondents a ``majority opinion that it is more important to teach the skills to use computational tools rather than the programming skills required to develop such tools, although a significant fraction feel that both are equally important.'' They particularly found that ``the skill to write code from scratch is not necessarily a crucial one'' \cite{enrique2018computational}, as is often advised within the community of computational physics educators \cite{weatherford2013mwp,oleynik2019scientific}. 

Interviewing industry professionals, academic faculty, and recent graduates who frequently engage in computational work, Caballero organized a set of necessary skills under the categories of knowledge of specific physics and computational concepts (iterative concepts, well-established numerical methods, and newer concepts for data science), use of specific computational practices (planning, modeling, programming, debugging, analysis, and visualization), computational ways of thinking (computational thinking, thinking iteratively, thinking discretely), ways of connecting mathematics, physics, and computational ideas (moving fluidly between mathematical, physical, and computational ideas, and seeing how each affects the other), meta-knowledge about computation (what computing can do and what tools are appropriate for certain tasks), and developing professional practices (documenting code and using version control) \cite{caballero15skills}.

Zwickl et al  \cite{zwickl17characterizing} highlight complementing analytical math to ``gain an early-stage understanding of a problem'' with using computational math to ``model via computation and simulations.'' In the analytical mathematics work, they emphasize understanding background concepts, modeling systems through equations, exploring simplified models, and understanding the impact of different parameters. In computational modeling, they emphasize extending simpler analytical models, understanding the impact of different parameters, visualizing data and representing models, analyzing and interpreting data, carrying out numerical calculations, and interfacing with data acquisition and databases. 

These studies summarize the consensus among STEM professionals of which computational practices are important in the physics and STEM communities' sense of joint enterprise. These are the scientific practices that form ``a set of activities that are embedded within and work toward the aims of a scientific community'' \cite{quan2018interactions}. Therefore, the COP framework can help instructors establish learning objectives for a computationally integrated physics curriculum that creates a local academic community that aligns with the global professional community. This alignment is part of the instructor's role as a ``broker,'' \cite{irving2014conditions}, and helps students chart appropriate computation-informed trajectories as they navigate the next stage of their careers within the global community. In this work, we are interested in investigating how students appropriate these computational practices as personally important.

\subsection{\label{sec:needs}Needs for Development}

A growing number of studies \cite{Caballero11Fostering,Dwyer13Computational,odden19computational,Caballero14Model,martin2016undergraduate,joglekar2020normalizing,Burke17Developing,bott2020student,Hawkins17Examining,caballero2012implementing,gavrin2021preliminary,weber2021conceptual} have found that computationally integrated physics coursework can successfully develop students' computational skills and conceptual understanding, and that undergraduate students can be overall satisfied with computationally integrated physics education, even when students believe their knowledge is limited \cite{lunk2016attitudes}. 
However, this rich pedagogical approach still requires investigation and improvement. Student resistance has been identified as a common and significant obstacle in integrating computation \cite{leary2018difficulties}, and not all students come to see computation as a helpful learning experience \cite{Hawkins17Examining}. Because computation is a novel activity for many students, and many students find programming intimidating \cite{lunk2016attitudes}, student motivation \cite{keller1987development} plays a key role in integrating computation into the physics curriculum. In the language of Communities of Practice, students need to perceive computation as part of their community's sense of joint enterprise and as an accessible set of practices they can successfully acquire along their unique trajectories. 

Students' ideas about computation develop through various stages of incompleteness \cite{Dwyer13Computational}, and, in some instances, students develop an incomplete skill set. For example, they might demonstrate that they have learned how to develop code but not to interpret its results \cite{Caballero14Model}. Similarly, Caballero, Kohlmyer, and Schatz described their students' difficulties with updating variables as ``highlight[ing] the fragility of their computational knowledge'' \cite{caballero2012implementing}. Some students' perceptions of computation's benefits focus on a subset of what computation offers \cite{bott2020student,lane2021comparison}. These ``not quite there yet'' stages of incompleteness represent progress along an inbound trajectory within the physics community. The adoption of computational practices for physics-relevant goals is part of navigating an inbound trajectory.

Anecdotally, we have observed and heard from other educators that students' perceptions of computation are not always positive, and that students can sometimes see computation as intimidating, irrelevant, or simply the specialized interest of a minority of physicists. A concern often voiced among physics educators who use computation is that, while students might be \textit{exposed} to computational methods and skills, they don't necessarily \textit{adopt} them as personally important, or see them as a \textit{normal} part of the physics community of practice \cite{joglekar2020normalizing,gavrin2021preliminary}. As a result, students might not transfer computational knowledge to a different domain or a new task \cite{caballero2012implementing}. In the language of Communities of Practice, there is a misalignment between students' perceptions of computation and the role of computation in the physics community's sense of joint enterprise. 

\begin{figure}[tb]
\includegraphics[width=0.45\textwidth]{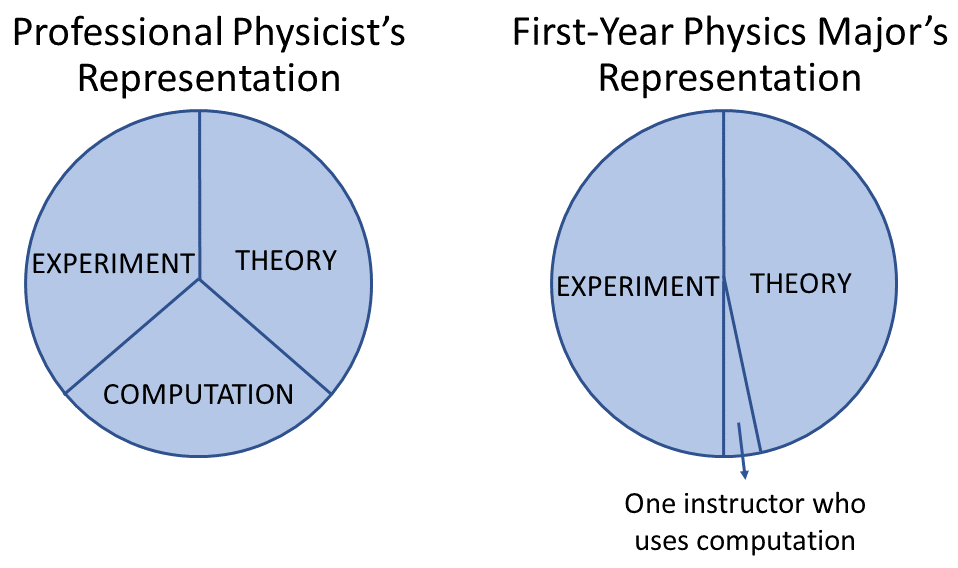}
\caption{\label{fig:ComputationPieCharts} Pie charts representing possible internal representations of physics practices that a professional physicist (left) and first-year physics major (right) might adopt based on their differences in experience. A professional is likely to give comparable space to computation, experiment, and theory, while a first-year major might have limited experience with computation, leading to different expectations of the practices used by the physics community.}
\end{figure}

We illustrate this difference in Figure \ref{fig:ComputationPieCharts} with two pie charts depicting possible representations of the practices employed by the physics community. The pie chart on the left depicts a representation a professional physicist is likely to adopt: The categories of experiment, theory, and computation occupy roughly equal prominence in the types of practices that physicists use. The pie chart on the right depicts a representation that many students seem to adopt: Physics practice is dominated by experiment and theory, with computation used by only a small fraction of the community.

This misalignment between experts' practices and students' perceptions can hinder students' progress along an inbound trajectory. For example, consider the student who learns how to carry out numerical integration in a computationally integrated upper-division electromagnetic theory course. The student might use a numerical integration code to evaluate the electric potential or electric field for dozens of charge distributions. Now suppose the next semester begins and the student is enrolled in a quantum mechanics course where computation is not integrated. The student \textit{could} use the numerical integration practices they learned in E\&M to evaluate the many integrals encountered in quantum mechanics, but only if the student perceives numerical integration as personally helpful and a valid approach in this new context. They need to transfer the usefulness of computation from the community of their E\&M class to the community of their quantum mechanics class, knowing that computation is valued in the global physics community. Such an extended application would be evidence of their progress along an inbound trajectory. If, instead, they relegate numerical integration to a course-specific requirement, they likely will not activate this knowledge \cite{krawitz2018activation,hofer2018enhancing,national2018people} to use in the new context \cite{caballero2012implementing}, stalling their progress along an inbound trajectory. Considering the COP framework can help address such transfer issues by maintaining the link between local academic activities and global professional practices \cite{wenger1999communities,irving2020communities}.

To explore this misalignment between novice perceptions and professional physics practices, we examine how students perceive computation as a physics practice \textit{after} encountering it in their coursework. This goal emphasizes the persistence of students' perceptions of computation after they are no longer required to engage in it. This persistence impacts the likelihood that a student will transfer computational practices to a new context \cite{caballero2012implementing}, which is a key step in progressing along their inbound trajectory. In this exploration, we employ the idea of a mental model to consider \textit{students' mental models of the physics COP} (which we call a COP-Model). We believe the COP-Model construct is useful beyond computation (and beyond physics), so we begin the next section by describing this construct in general terms, and then apply it to our considerations of computational practices in physics education.

\section{\label{sec:COPModel}Student Mental Models of the Physics Community of Practice} 

In this section, we discuss the need to attend to students' mental models of the COP they are navigating, particularly in the context of the physics COP. After introducing terminology from the mental models framework, we explore how a student's mental model guides them as they navigate a COP, and discuss how these models are important to the student experience.

\subsection{Mental Models\label{subsubsec:mm}}

The mental models framework describes how learners use internal representations to guide their reasoning and form expectations based on experiences \cite{greca2002mental}. Mental models are ``abstract representations that store the spatial, physical and conceptual features of experiences'' \cite{rapp2005mental}, or ``systematically constructed representations of physical systems, used to describe, represent, and explain the mechanisms underlying physical phenomenon'' \cite{louca2011objects}. Mental models are built out of predictions and explanations for the reality they imitate \cite{thacker2019visualizing,greca2001mental}. Learners use mental models to facilitate ``retrieval in the service of problem solving, inference generation and decision making” \cite{rapp2005mental} and to ``ground abstract scientific ideas'' \cite{thacker2019visualizing} that they can then analyze mentally and create new inferences and associations \cite{schwartz2006spatial}.

The formation of mental models lies at the heart of much of the learning process. Mental models are formed from the learner's experiences \cite{thacker2019visualizing,greca2001mental}, and, as such, mental models are unique across individuals and dynamic over time \cite{wade2018modeling}. Model-based learning can be understood to start with learners' pre-existing models (preconceptions) and designed to reach a target model (learning objectives) at the end of a series of intermediate models (partial understanding) \cite{clement2000model}. Louca et al consider mental models to have five elements \cite{louca2011objects}: 1. physical objects, 2. physical entities, 3. object behaviors, 4, interactions among objects, entities, and behaviors, and 5. accuracy of the model's descriptions. 




\subsection{The COP-Model: A Map for Career Navigation}

\begin{figure*}[tb]
\includegraphics[width=\textwidth]{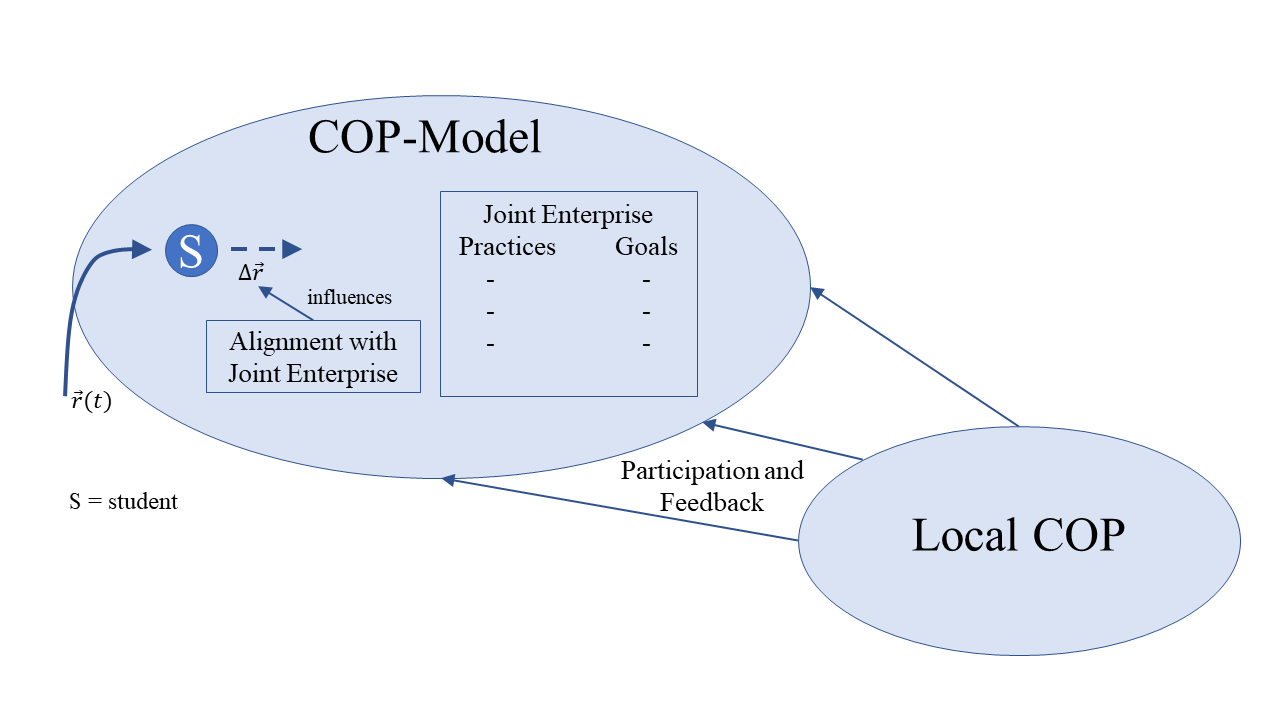}
\caption{\label{fig:COP-Model} The construct of a COP-Model. The student's mental model includes their trajectory history ($\vec{r}(t)$), their next step ($\Delta \vec{r}$), and their understanding of the community's sense of joint enterprise. Their next step is informed by their current sense of alignment with the sense of joint enterprise. All these elements are shaped by the legitimate peripheral participation in the local COP. We apply this construct specifically to the use of computation in physics education, but believe it can offer insights into all STEM communities.}
\end{figure*}

We suggest that an important factor in a learner's pursuit of any STEM career (not just physics) is their \textit{mental model of the Community of Practice} (which we call a COP-Model) relevant to that career. Attending to these models is important, as students' dispositions toward STEM are heavily influenced by the professional and personal authenticity of their STEM learning experiences \cite{knezek2013impact,christensen2013contrasts,shin2015changes,means2021cultivating}, and these dispositions influence the type of trajectory a student pursues within a STEM community, thereby impacting representation within the STEM community. In the classroom context, a student's inbound trajectory begins with authentic learning experiences that promote a COP-Model that aligns with the global COP as it exists in reality. We illustrate our construct of the COP-Model in Figure \ref{fig:COP-Model}, which we unpack throughout this section. 

In a COP-Model, the system being represented is the global professional COP that a learner is navigating as a new member based on their interactions with their local academic COP. Using the list of elements from \cite{louca2011objects}, we identify this mental model as consisting of...
\begin{enumerate}
    \item \textit{Objects}: individuals and institutions such as practitioners and departments. This set of objects includes the learner, other members of the local academic community (such as classmates and professors), and members of the global professional community (such as conference presenters or published authors).
    \item \textit{Entities}: qualities of those individuals and institutions that describe their membership within the community. These qualities include the learner's expectations of participating in the COP, confidence, position, and trajectory. 
    
    \item \textit{Behaviors}: actions and practices that the individuals and institutions engage in. This set of behaviors includes the learner's legitimate peripheral participation and examples they observe carried out by central community members. 
    \item \textit{Interactions}: the sense of joint enterprise that guides the individuals and institutions, establishes standards for qualities, and mediates actions and practices. 
    
    \item \textit{Accuracy}: the alignment between the learner's COP-Model and the COP-in-reality. 
\end{enumerate}

We further describe a COP-Model as follows.


\subsubsection{A learner's COP-Model includes the goals and practices that make up their understanding of the COP's sense of joint enterprise.}

The COP's sense of joint enterprise is represented as interactions among the model's objects, entities, and behaviors. This representation (a list of goals and practices in Figure \ref{fig:COP-Model}) helps the learner answer questions like, ``What common goals guide individuals and institutions to perform certain practices?'' or, ``What go-to practices do members of this community use to reach their common goals?'' or, ``What counts as doing physics?'' We emphasized earlier that inbound trajectories are characterized by the novice's adoption of the COP's goals and practices \cite{lave1991situating,irving2020communities,barab2000practice,irving2016identity}. We now emphasize that, before a novice can adopt those goals and practices, they must be represented in the novice's mental model of the COP (list of goals and practices in Figure \ref{fig:COP-Model}). Conversely, a novice won't adopt a goal or practice they don't first see in their COP-Model. As mentioned earlier, the novice's perceptions and values (such as their personal answer to the question ``What counts as doing physics?'') must be shaped to align with the COP-in-reality \cite{close2016becoming,irving2014conditions}; these perceptions and values are held in the novice's COP-Model. 

A newcomer to a COP is likely unfamiliar with these goals and practices and has no representation (or a misaligned representation), while an established member near the COP's center is likely to have a well aligned representation of these goals and practices. Many physics educators observe this phenomenon anecdotally in introductory mechanics courses when students think that practicing physicists spend their time solving free-body diagrams or studying the motion of projectiles. Such a mental representation of the physics community's sense of joint enterprise lacks extrapolation to more modern topics. Or, as discussed earlier, a student in the introductory context might not be aware of the significant role that computation plays in physics practice, such that their representation of the sense of joint enterprise would not include computational practices. 

We see the COP-Model as what Irving, McPadden, and Caballero are referring to when they write that the identity that a learner develops in the local academic COP ``give[s] the participant a sense of how their practices and participation fit within a broader context'' \cite{irving2020communities}. 





\subsubsection{A learner's COP-Model includes their sense of membership within the COP.}

When studying students' participation in informal physics outreach, Fracchiolla, Prefontaine, and Hinko employed an operationalized Communities of Practice framework that ``allows us to sense where the university students see themselves within the community of practice and to learn what aspects of that community impact their involvement'' \cite{fracchiolla2020community}. Similarly, we see the COP-Model as a map that guides the learner along their trajectory ($\vec{r}(t)$ and $\Delta \vec{r}$ in Figure \ref{fig:COP-Model}) within the COP. This perceived position and trajectory are informed by the learner's confidence in their ability to contribute to the COP's goals using the COP's practices. 
This sense of position helps them answer the questions, ``Where do I fit in this community?'' and, ``What does this community think of me?'' \cite{kelly2016social}. For novices on an inbound trajectory, increased confidence scales inversely with their perceived distance $|\vec{r}(t)|$ from the center of the COP. This spatial element depicts the relative position of the learner within the community, and the learner's next step along their trajectory. 

A learner's perceived trajectory gives them an indication of their progress toward or away from the center of a COP. This perceived trajectory helps the learner negotiate their experience of self in comparison to the COP and answer the question, ``To what degree is this community's joint enterprise \textit{my} enterprise?'' These considerations inform the learner's trajectory (continuing toward the center or periphery, changing direction, changing speed) and career decisions (the degree to which their future career is connected to this community). This sense of progress is informed by their sense of how the COP's goals align with their own interests \cite{diekman2011malleability} and how the COP's practices align with their own competencies. We expect that an inbound trajectory is accompanied by an increasing sense of alignment with the COP
.


\subsubsection{A learner develops their COP-Model in response to 
legitimate peripheral participation and feedback.}

Mental models are built from experiences, usually in an attempt to make the model better align with reality and thereby improve its usefulness as a reasoning guide. A student progressively builds their COP-Model from 
legitimate peripheral participation \cite{franz2016experiences,boylan2016deepening,irving2020communities} and the resulting feedback (arrows between the two ellipses in Figure \ref{fig:COP-Model})
. As noted earlier, ``Identity as membership in a COP connects identity with forms of competence'' \cite{irving2020communities}. The learner's interactions with the local COP develop and demonstrate the learner's competencies with the COP's practices through requirements (what the community expects) and resolution (the community's evaluation). Similarly, 
legitimate peripheral participation guides novices in appropriating the practices of the community as they move along an inbound trajectory \cite{barab2000practice,irving2016identity,irving2020communities}.

The COP's culture, goals, and approaches determine the nature and expectations of legitimate peripheral participation, such that this participation refines the sense of joint enterprise represented in their COP-Model (list of goals and practices in Figure \ref{fig:COP-Model}). In the learner's COP-Model, the competencies that legitimate peripheral participation demonstrates are qualities (entities) that describe the learner (an object) and help evaluate their alignment with the COP (their ``understanding of self within [the] community'' \cite{irving2020communities}) and identify their position and trajectory. From this perspective, one goal of legitimate peripheral participation is for learners to develop a COP-Model that is aligned enough with reality to guide their interactions with the COP and form expectations as they navigate the COP.

Misalignment between students' COP-Models and the corresponding COP-in-reality has been observed in STEM education research. For example, Franz-Odendaal, et al found that ``Grade 7 students do not grasp the importance of science/math requirements for future STEM careers'' \cite{franz2016experiences}. ``Grasping the importance'' of competencies expected in a career is not a matter of understanding or even developing those competencies, but perceiving them as a valuable requisite of one's career goals. This sense of value is represented in the students' COP-Models.

Similarly, in the context of information technology, Agosto, Gasson, and Atwood argue that one reason female students divert from IT careers is that they do not perceive that an IT career can overlap with their goal of solving problems \cite{agosto2008changing}. In the broader context of STEM, Diekman et al argue that women opt out of STEM careers because they see STEM fields at odds with fulfilling communal goals \cite{diekman2010seeking}. These expectations are misaligned from how IT professionals would describe problem-solving as a key element of their jobs and how STEM professionals would describe their careers as fulfilling communal goals. Helping female students incorporate this ``goal congruity'' \cite{diekman2011malleability} into their COP-Models is an important goal in reforming STEM education. 



\subsubsection{A learner's COP-Model enables them to extrapolate their experience of a local COP to an understanding of the global COP.}

We see this purpose as especially salient for undergraduate students, whose experience of the COP-in-reality is limited to the context of their local COP within an academic program or research group. They must extrapolate this experience to a model of the global professional COP \cite{quan2018interactions}. 

As a fictional analogy, consider the 2005 film \textit{Robots}. The protagonist, Robbie, has grown up dreaming of a life in Robot City, working for Bigweld Industries, where he believes he will be welcomed to bring new ideas, receive support in actualizing those ideas, and contribute to the betterment of robot society. This is Robbie's model of the Bigweld Industries community of practice, which he developed from representations of the company portrayed on television during his childhood. About halfway through the movie, Robbie discovers that the company has been taken over by new leadership that has changed the community into a hostile and classist environment, not even allowing him to enter the company grounds. The community's sense of joint enterprise has changed, causing severe cognitive dissonance for Robbie. On a phone call with his father back home, he despondently relays, ``It's not like we thought.'' The COP-in-reality, he has discovered, is drastically different than his COP-Model.

\subsubsection{Comparing models of different COPs helps a learner develop and negotiate their nexus of multimembership.}

No student is a member of the physics community only, a principle which Wenger refers to as a nexus of multimembership \cite{wenger1999communities}. The models a physics student develops of the various communities to which they belong (family, neighborhood, academic, athletic, professional, religious, etc.) helps them understand how these communities intersect and support each other or stand disparately from each other. 
These mental models allow the learner to easily juxtapose their communities' goals and practices, compare their sense of membership within each community, 
and negotiate their responsibilities toward each community. Such comparison might play a crucial role in the learner's decision of whether to persist along an inbound trajectory in a given community (``Is this new community compatible with my existing communities?''), a process which has direct impacts on the community's representation (``Why is no one from that community centrally involved in this community?'') \cite{cheryan2011female,diekman2010seeking,diekman2011malleability,moshfeghyeganeh2021effect}. From this perspective, representation becomes a result of communities comfortably sharing central members based on the development of compatible COP-Models. Once multimembership is established, the interplay between models of different communities enables the shared member to transfer ideas from one community to another, helping to develop the sense of joint enterprise.

\subsection{\label{COP-Model-teaching}COP-Models in the Classroom}

It is important that educators attend to how well a student's COP-Model reflects reality, since (1) a student who professes interest in a STEM career but has unrealistic expectations of that career is unlikely to persist \cite{rodriguez2015developing} and (2) a student who declines to pursue a STEM field but whose COP-Model does not include appealing facets of that field is missing out on a potentially rewarding career and might unnecessarily opt out of the field's employment pool \cite{aschbacher2010science}. The formation of students' COP-Models is part of the teacher's role as a ``broker,'' shaping their local academic community to better match the global professional community \cite{irving2014conditions}. We argue that educators would do well to attend to the development of students' COP-Models as part of the scaffolding process in a course, such that students can begin to chart their own inbound trajectories rather than following the trajectories explicitly laid out by their instructors and mentors \cite{rodriguez2015developing}. Such attention is likely to affirm a greater diversity of student trajectories into the STEM community and thereby support representation within that community.

In particular, we believe this construct helps to inform several important learning objectives:
\begin{itemize}
\item Students will formulate reasonable career goals in relation to STEM interests and expectations.
\item Students will develop a positive perception of themselves and their classmates as ``STEM people'' \cite{Lock2013Physics}.
\item Students will practice activities relevant to STEM professions.
\end{itemize}

Additionally, students' COP-Models can impact persistence and diversity within physics professions, since ``What type of people are included in the physics community?'' and ``Do others see me as a physicist?'' is addressed in a student's COP-Model \cite{Lock2013Physics,hazari2020context}. Educators and mentors might be able to proactively develop learners' COP-Models using metacognitive tools demonstrated to help students develop and use mental models \cite{wade2018modeling}. 

We think the construct of a COP-Model warrants further investigation, and that it could be used to design assessments and interventions to help educators more directly develop students' perceptions of physics as a professional community. With this construct established, we restate our overarching question (``How do students develop expert-like perceptions of computation as a practice within the professional physics community?'') more specifically as, ``How is computation represented in students' mental models of the physics community of practice after they complete a computationally integrated physics course?''

\section{\label{sec:ContextMethods}Context and Methodology} 

Based on this conceptual framework, we conducted semistrcutred interviews of students to answer the above question. We also interviewed these students' instructors to foreground their perceptions in the community of practice these instructors created across their computationally integrated physics courses. In this section, we review the academic context for our study, expand this primary question into three research questions, and outline our study procedures.

\subsection{Academic Context}

We investigated our primary question in the context of the physics program at a mid-sized primarily undergraduate regional state university. This program graduates 10-20 physics bachelors each year, emphasizes research opportunities for undergraduates, and has 9 tenure-track or tenured faculty, 5 non-tenure track faculty, and 6 visiting full-time faculty. In fall 2020, three faculty members (including this paper's first author) integrated regularly occurring computational assignments into three upper-division physics courses (Astrophysics I, Mathematical Physics, and Modern Physics). This implementation of computation places this department among the 52\% of departments in the USA reported to have at least 1 faculty member teaching computation in an advanced-level physics course, and the 39\% of departments whose faculty use computational homework \cite{caballero2018prevalence}.

A handful of students were enrolled in more than one of these courses, such that they experienced computation in multiple physics courses in the same semester. The students in these courses had varying degrees of prior experience with programming. Some had taken an introductory programming course offered by the computing sciences department, while most had never encountered programming beyond using spreadsheets in introductory physics labs. 

The instructors had all participated in computation-based professional development through the Partnership for Integration of Computation into Undergraduate Physics \cite{PICUP:main} and designed their computational activities to orient students to relevant programming tasks (such as arrays, control statements, and loops) before introducing physics applications. The three instructors discussed their plans and progress throughout the semester. Following departmental norms, computational activities took the form of guided minimally working programs \cite{oleynik2019scientific,lunk2012framework,weatherford12student,weatherford11student} 
written in Jupyter notebooks (Python) on a university server. Therefore, students enrolled in more than one of these courses had similar learning experiences across their fall 2020 coursework. Because this common set of students had similar experiences in classes that were taught with similar frameworks, we define the community of practice for our study as the students and instructors within these courses, of which the interviewed students and instructors are a sample. The practices of this community, broadly, fall under the use of computation to explore and understand physics concepts. We further consider this local academic community as an expression of the global professional physics community, which also uses these practices.

\subsection{Research Questions}

We decided to study how these students' COP-Models represent computation and their trajectories within the global physics COP. We conducted this study \textit{after} the students completed their fall 2020 coursework involving computation because we wanted to examine the persistence of their experience with computation. This study therefore complements Hawkins et al \cite{Hawkins17Examining}, who interviewed students about their perceptions of using computation shortly after completing the first three weeks of a computationally integrated physics course. With this context established, we expanded our overarching question (``How is computation represented in students' mental models of the physics community of practice after they complete a course involving computation?'') into the following research questions:

\begin{enumerate}

    \item How did the instructors' background with computation inform how they established the local COP in these courses? This question helps us foreground the students' perceptions in the context of the local COP.
    \item How does computation feature in each student's COP-Model? Do they see computation as normal (embedded within the ensemble of activities that the physics community uses) \cite{quan2018interactions,Ford06,joglekar2020normalizing,gavrin2021preliminary}, as valuable (offering a useful contribution to their knowledge) \cite{quan2018interactions,Berland16}, and connected to other physics practices \cite{quan2018interactions,Ford06}?
    \item Where does each student see themselves in the professional physics COP, as assessed by their use of computation? Do they feel confident in their ability to use computation? Do they see further opportunities to use computational practices? We base this question partly on Caballero, Kohlmyer, and Schatz's observation that ``an evaluation of transfer [of computational knowledge] would require that students apply these computational skills to a different domain... or a different task'' \cite{caballero2012implementing}. To our knowledge, this evaluation of transfer has not been carried out in a formal PER study.

\end{enumerate}

\subsection{Study Design and Participants}

We explored these questions with semistructured interviews in the spring 2021 semester, 2-3 months after the students completed these computationally integrated courses. We designed this study as exploratory, obtaining descriptive data in a case-based structure designed to elucidate possible avenues for student learning \cite{CaseBased,quan2018interactions}. We acknowledge that this timeline places the computationally integrated courses and the interviews in the midst of the COVID-19 pandemic. As such, the three courses featured split remote/on-site teaching formats, and the interviews took place over video conference. 

Students and instructors were recruited in January through March 2021 from the fall 2020 class rosters, with interviews conducted in February through April 2021. Five students (Chrissie, Guy, Harrison, Jose, and Paul) and all three instructors (Charles, Lance, and Marlee) agreed to participate. The students' course enrollments and prior experiences with programming are listed in Table \ref{table:students}. As one of these instructors is the first author on this paper, all instructor interviews were conducted by the second author. 

The student interview questions (Section \ref{sec:protocols}) were designed to prompt students to reflect on their computationally integrated courses chronologically, helping us understand their perspectives of their trajectories into the COP. 

We used Zoom to host, record, and transcribe these interivews. The authors shared responsibility for analyzing interview transcripts. The first author reviewed and coded the other instructors' interviews first and then used his own transcript primarily to confirm themes already identified. 

\begin{table}[b]
  \caption{Summary of student interview subjects. \label{table:students}}
  \begin{ruledtabular}
    \begin{tabular}{|p{0.07\textwidth} | p{0.19\textwidth} | p{0.19\textwidth}|}
    Student & Physics Enrollments & Programming Background\\ \hline
    Chrissie & Math. Phys., Mod. Phys. & Prog. I \& II \\
    Guy      & Astro. I, Math. Phys. & Prog. I \\
    Harrison & Math. Phys., Mod. Phys. & Prog. I \\
    Jose     & Math. Phys., Mod. Phys. & None \\
    Paul     & Math. Phys., Mod. Phys. & Prog. I \& II \\
    \end{tabular}
  \end{ruledtabular}
\end{table}

We sent student subjects a pre-interview survey to ask about their prior experiences with computation, confirm which computationally integrated physics courses they participated in, solicit their career goals, administer the Computational Thinking Attitudes Survey (CTAS) \cite{megowan-romanowicz_2020}, and collect a pseudonym for them to be referred to during the interview. The CTAS assesses students' attitudes toward the use of computation in physics, and was included to complement their interviews. Instructors were sent a pre-interview survey to collect a pseudonym.

The interviews followed a semistructured protocol \cite{drever1995using} in which we outlined a set of general questions and established space for follow-up questions and clarifying statements based on the subjects' answers. The interview protocols and pre-interview survey are located in Section \ref{sec:protocols}.

We analyzed interview transcripts using the method of constant comparison \cite{taylor1984introduction}, in which each author independently coded the transcripts to identify emergent themes that address our research questions. We revisited our theme definitions and coding criteria throughout the process and compared our themes and coding incidents after completing the full set of transcripts. This comparison produced an agreed-upon set of themes and coding criteria, which we used to review the transcripts once more to finalize the coding and select excerpts for this paper.

\section{\label{sec:Interviews}Interviews}

This section summarizes the contents of our student and instructor interviews. In Subsection \ref{subsec:Instructors}, we foreground the students' responses with a description of the community of practice these instructors sought to create within and across their courses. In Subsection \ref{subsec:CTAS}, we present an overview of the students' attitudes toward computation as reflected in the Computational Thinking Attitudes Survey. In Subsection \ref{subsec:Students}, we review highlights from each student's interview, building a profile of each student and how computation is represented in their physics COP-Model. Finally, in Subsection \ref{subsec:Themes}, we discuss themes that emerged across the students' interviews.

\subsection{\label{subsec:Instructors}Community Context: Insights from Instructor Interviews}

Young et al observed that faculty's personal experience with computation and beliefs about computation impact their decision to integrate computation \cite{young2019identifying}, so we interviewed the instructors of these three courses about their use of computation in teaching and research. These interviews yield insight into the community of practice created across these courses. We believe this COP was reinforced by coordination among the instructors and many students being enrolled in two or more of these courses. These interviews yielded additional insights into the instructors' pedagogical decisions, which we will explore in future work.

These instructors identified professional authenticity and student career readiness as important reasons for integrating computation. They frequently referred to the demands for computational skills in graduate study and industry. Out of this motivation, these instructors based their integration of computation on their research experiences. All three instructors described using computation in their research activities, and they made comparisons between the practices they use frequently in their research and the practices they incorporated into their courses' computational activities. For example, Charles described his research as primarily consisting of analyzing large data sets to develop models, and this process formed the structure of many of his computational activities.

These instructors also based their integration of computation on the support computation offers for student learning objectives. They each described computational activities as supporting student learning (alongside analytical and experimental activities). However, they reported varying degrees of pedagogical competence in designing computationally integrated coursework.

These instructors believe that computation offers a breadth of benefits to learning and doing physics, such as visualization, efficiency, accuracy, data analysis, and sense-making. These benefits are valued by the global professional physics community \cite{Graves20Hitting,Burke17Developing,thornton2009computational,enrique2018computational,caballero15skills,zwickl17characterizing}, and these instructors designed their computationally integrated physics courses to highlight these benefits. For example, Lance and Marlee each directed students to complete problems analytically and computationally to contrast the differences between each approach.

Finally, these instructors identified students' lack of programming background as a barrier to learning. Marlee and Charles seemed particularly concerned about initiating students into the Python programming language, and all three instructors expressed concern for students' confidence. This shared concern motivated all three instructors to use minimally working programs \cite{weatherford2013mwp} in their computational activities. 

Based on these observations, we conclude that these instructors sought to create a local academic community of practice within and across their courses. They viewed their courses as a local embodiment of the global professional physics community. Within this view, they supported their students' transition from the local academic community to the global professional community by creating opportunities for legitimate peripheral participation.

\subsection{\label{subsec:CTAS}Student Pre-Interview Survey}

Prior to the interview, students completed a survey (presented in Section \ref{subsec:survey}) about their experiences with computation and the computationally integrated physics courses they completed in fall 2020 (Table \ref{table:students}). This survey also asked students to rate their agreement with a series of statements from the Computational Thinking Attitudes Survey (CTAS) \cite{megowan-romanowicz_2020}. Their responses to these statements are presented in Table \ref{table:CTAS}.

\begin{table*}[hbp]
  \caption{Student interview subjects' responses to the Computational Thinking Attitudes Survey (1 = Strongly Disagree, 2 = Disagree, 3 = Neutral, 4 = Agree, 5 = Strongly Agree) \cite{megowan-romanowicz_2020}. \label{table:CTAS}}
  \begin{ruledtabular}
    \begin{tabular}{p{0.5\textwidth}ccccc}
    \centering Student & Chrissie & Guy & Harrison & Jose & Paul \\
    \hline
    1.1 Generally, I feel comfortable working with computers. & 4 & 5 & 5 & 3 & 5 \\ \hline
    1.2 Generally, I am competent using computational tools (such as spreadsheets, graphing programs, simulations and writing code). & 4 & 4 & 4 & 2 & 4 \\ \hline
    1.3 I regularly use computation tools  (such as spreadsheets, graphing programs, simulations and writing code). & 3 & 4 & 4 & 2 & 3 \\ \hline
    1.4 I regularly collect data when problem solving. & 4 & 5 & 3 & 4 & 3 \\ \hline
    1.5 I often analyze data when problem solving. & 4 & 5 & 4 & 2 & 3 \\ \hline
    1.6 Visualizing data is important in understanding a concept. & 5 & 5 & 4 & 5 & 4 \\ \hline
    1.7 Having background knowledge in computers ( and computational tools)  is valuable. & 5 & 5 & 4 & 5 & 5 \\ \hline
    2.1 I am able to deal with open-ended problems that are related to computer skills. & 5 & 4 & 4 & 3 & 4 \\ \hline
    2.2 The challenge of solving problems with computers appeals to me. & 5 & 4 & 4 & 4 & 4 \\ \hline
    2.3 I can break down larger problems into smaller problems. & 4 & 4 & 4 & 4 & 4 \\ \hline
    2.4 I am comfortable learning new computing concepts. & 5 & 4 & 4 & 2 & 5 \\ \hline
    2.5 I have an idea of what to do when I see an error message. & 4 & 4 & 4 & 4 & 4 \\ \hline
    2.6 Error messages provide useful feedback when I am coding (writing a computer program). & 4 & 2 & 4 & 4 & 4 \\ \hline
    3.1 I enjoy doing computer programming when learning physics. & 5 & 4 & 4 & 2 & 4 \\ \hline
    3.2 I like to use or write computer simulations to learn physics. & 5 & 4 & 4 & 4 & 3 \\ \hline
    3.3 Using computational tools helps me to understand physics concepts in a different way. & 5 & 5 & 4 & 4 & 4 \\ \hline
    3.4 Learning to create  computer simulations in physics will make me better in math. & 5 & 4 & 3 & 4 & 5 \\ \hline
    3.5 Writing a computer program helps me to explain my understanding of physics to others. & 5 & 5 & 3 & 4 & 3 \\ \hline
    3.6 I don't need to know how to write computer programs to understand physics and solve physics problems. & 3 & 4 & 4 & 4 & 3 \\ \hline
    3.7 It takes longer to solve physics problems using programming. & 2 & 3 & 3 & 4 & 3 \\ \hline
    3.8 I understand physics concepts better when I can construct and use computational models that illustrate these concepts. & 5 & 5 & 3 & 3 & 3 \\ \hline
    4.1 Overall, I am motivated to learn more about how to use computers to learn physics. & 5 & 4 & 4 & 4 & 5 \\ \hline
    4.2 Overall, I am motivated to develop and use computational models of physical phenomena. & 5 & 4 & 4 & 4 & 4 \\ \hline
    4.3 Overall, I am motivated to use data when I'm problem solving. & 5 & 5 & 4 & 4 & 4 \\ \hline
    4.4 Overall, I am motivated to use computation tools  (such as spreadsheets, graphing programs, simulations and writing code). & 5 & 5 & 4 & 4 & 4 \\ \hline
    4.5 Overall, I am motivated to solve problems by using computer applications. & 5 & 4 & 4 & 4 & 4 \\ \hline
    4.6 Overall, I am motivated to analyze and interpret data. & 4 & 4 & 4 & 5 & 4 \\ \hline
    4.7 Overall, I am motivated to use computer programs  to understand physical systems. & 5 & 4 & 4 & 5 & 4 \\
    \end{tabular}
  \end{ruledtabular}
\end{table*}

The students' responses to the CTAS depict an overall expert-like disposition toward computation after completing computationally integrated physics coursework. They tend to value and feel motivated to use computation-centered practices, as seen in statements 1.4-1.7, 3.1-3.5, 3.8, and 4.1-7. However, they provide mixed responses to statements about their confidence with computation, as seen in statements 1.1, 1.2, 2.1, and 2.4-2.6. Jose in particular seems to lack confidence in many computation-related skills. We find these trends are also present in the student interviews.

\subsection{\label{subsec:Students}Student Interviews}

The following subsubsections summarize the insights we gained from each student's interview. We build a profile of each student and how their COP-Model represents computation.

\subsubsection{Chrissie: Confident with Minor Misalignment}

Chrissie came to see computation as a valuable tool in experimental physics research, and in studying physical systems that are otherwise mathematically inaccessible. This aspect of her COP-Model aligns with the professional physics community \cite{chonacky2008integrating,weller2018investigating,Serbanescu11Putting,chabay08computational,young2019identifying}. She particularly values computer visualization as a means to learn new physics concepts. Here, she describes the example of the infinite square well problem from Modern Physics: 

\quotepar{Chrissie}{Through the code, you can kind of see see like through a graph that this is what happens at this energy, and this is why this other thing happens related to energy levels... And you could instantly just test it and see what what happened, and you can look at the graph and see, ``Oh, so this goes towards this boundary or does it go towards this boundary''... It's different from just looking at the integrals involved, because then you get a number at the end. It's like, you don't really know that means.}

This use of computation aligns with the instructors' purposes for integrating computation into their courses, and with the global professional physics community's use of computation. 

Along with this alignment, Chrissie seemed particularly confident in her ability to use computation, and ties this confidence to her prior programming experience. However, she did report struggling with creating new code instead of adapting or extending code from a MWP. As noted earlier, coding from scratch is not highly prioritized in the global professional COP \cite{weatherford2013mwp,oleynik2019scientific}, as is reflected in the instructors' use of MWPs. This is an example of misalignment between Chrissie's COP-Model and the COP-in-reality. However, this confidence-reducing misalignment does not seem to have negatively affected her overall perceptions of computation or physics.

Chrissie also seemed less confident in bridging from computation to sense-making.

\quotepar{Chrissie}{Since I've done a lot of code before it wasn't really that bad... I feel like the hardest part was just doing the math and physics... just interpreting the physics, in terms of code.}

Chrissie foresees an industry career heavily steeped in computation, and did not report being involved in any undergraduate research at the time of the interview. 

\subsubsection{Guy: A Misaligned Lack of Confidence}

Guy described computation as being useful in an experimental context, both to work with and present data, and highlighted the necessity of computation in a STEM career. He particularly mentioned the use of computation in processing large data sets, an activity that was frequently highlighted in his Astrophysics I course.

Guy described visualization as a helpful purpose of computation in Mathematical Physics: 

\quotepar{Guy}{[Computation] gave you a visualization of how the math worked from class, so we would use programming to see how the different formulas and the different equations came out and graphs and pictures and different sorts of computational methods to see different ways that they work and it just gave a different view instead of just having numbers in front of you. You had graphs and you can see what's happening versus just trusting that your numbers are correct.}

Throughout the interview, Guy expressed his concerns over his lack of confidence with programming.

\quotepar{Guy}{In the beginning [of the semester] I wasn't comfortable at all... I did gain comfort levels over time, but I still can't---if you told me to take a data set and plot a data set and show it with any kind of analyzing... the only way I could do it is to steal somebody's code from, you know, messing around on Google... It doesn't feel like I'm coding, it feels like I'm playing plagiarism...  Even after both those classes, I don't feel comfortable when it comes to, you know, if I had a blank Python notebook to start collecting data on that computational side of things.}

This interview segment reveals a misalignment with the practices of the physics COP: In his COP-Model, Guy believes that professional coders would not use (``steal,'' as he describes it) preexisting code from an internet search (``messing around on Google'') to solve a problem, but would start from a blank Python notebook. In contrast, searching for sample codes is exactly what expert programmers frequently do, rather than starting a program from scratch \cite{Graves20Hitting,weatherford2013mwp,oleynik2019scientific}. In other words, his lack of confidence stems from failing to meet expectations that the physics COP does not actually hold.

Another misalignment (on the part of the community, it seems) is found when Guy describes his prior programming experience:

\quotepar{Guy}{I learned about C++ in my first year, which I wish it was never C++, I wish it was Python, because Python is used by physicists in the real world. C++, I mean it's a nice course to learn about basic coding, but it completely feels different. It writes different than Python does.}

Physics educators who integrate computation often express that transferring learning between programming languages is a straightforward process; this does not seem to be the case for Guy. We recommend the community reexamine the challenges in this process.

However, despite the trepidation that Guy expresses during the interview, he concludes by describing his expectation to use computation frequently in his career. It seems that, in Guy's COP-Model, he is far from the center (perhaps farther than he is in reality) but maintaining a steady inbound trajectory.

\subsubsection{Harrison: Enthusiastic but Tentatively Inbound}

Harrison described computation as important in his future career as a science communicator. Harrison enthused about the accuracy and efficiency offered by computation, particularly as efficiency helped him learn.

\quotepar{Harrison}{It alleviates so many small errors that could have occurred. Now, there are small errors that can occur in programming. But I think those are far beyond, as far as magnitude goes, of going through these problem-solving skills with your hand. And being that it's a computer, it's cold calculations, it's done in an instant... It's one thing to say, ``Okay, I see this in handwritten work that would take me three years to finish,'' but Python can do it in like 15 seconds... Not just do one handwritten [problem] the whole year... Now if computation wasn't a thing, and I had to do this all by hand, it would be vastly different because I wouldn't have the time.}

Like Harrison, the professional physics community values using a computer's efficiency to allow the researcher to focus their efforts on sense-making instead of tedium. His COP-Model accurately represents this value.

Harrison also describes how he developed confidence at different rates in his two courses:

\quotepar{Harrison}{At the very first I was completely overwhelmed... I [became] very comfortable with Modern Physics after I finished the first wave function assignment. After that I understood what was happening... Mathematical Physics took me until about the last week of the semester to be comfortable with the code. That's not a complaint. I know that class is difficult, but it took a very heightened level of concentration to understand the code in that class.}

Unfortunately, Harrison seems to have lost some confidence after not continuing to engage in computation during spring 2021. This might indicate that he feels he has moved away from the COP's center in some respects.

\quotepar{Harrison}{Being that I haven't worked in a direct Python terminal in about two to three months, I probably won't be as comfortable as I was during finals... Not to say, ``Oh, I didn't learn anything. I'm not retaining anything.'' It's that computational physics is one of those things where it has to have your attention for a very long amount of time so that you can retain these concepts... I still understand the very building blocks of Python and I can look at a code and understand what it's doing, but as far as the intricate physics concepts involved, I would not be able to recite them.}

Also, in Harrison's COP-Model, there is a hierarchy within the community's practices: ``building blocks'' (which seem to be located in the periphery) and ``intricacies''  (which seem to be located in the center). Harrison indicated he was not participating in any research at the time of the interview, which might contribute to his sense of regression.

\subsubsection{Jose: Emerging Alignment with Mixed Confidence}

Jose contrasted the use of computation in the context of an experiment with the use of computational modeling, which he finds more interesting. It seems his COP-Model includes multiple avenues for using computation, but he has clearly identified which uses of computation align best with his interests. He also expressed the importance of using computation to create visualizations that help one understand the physics concepts at work in a problem. 

Like Harrison, Jose described a difference in the rate at which he developed confidence in his courses:

\quotepar{Jose}{In Modern Physics, I felt more comfortable because it was, like, kind of bits and pieces that we would have to put together of code, rather than like a whole program. And Math Physics, there are times when it would be like a kind of like a whole program that we may have to put together and that was a little more difficult at the beginning.}

These differences (``bits and pieces'' versus ``whole program'') can be thought of as two examples of legitimate participation (both use programming to understand physics) with differing degrees appropriateness for Jose's position at the periphery. This difference is reflected in his COP-Model as different levels of confidence.

Jose seems relatively confident when asked how comfortable he would feel returning to his courses' computational activities:

\quotepar{Jose}{I think I'd feel fine going back and working through them... I have been graded for those assignments, so I know that I did whatever properly. Yeah, so I think that's the big factor is now that I have feedback that everything was going well, so I was doing it correctly.}

When discussing how Modern Physics never required him to write a code from scratch, Jose approaches this idea with relative optimism:

\quotepar{Jose}{I think having like maybe one assignment near the end where it's like a more simple problem that we would have to kind of create a code from scratch would be really fun to do and helpful and just kind of like piecing together what all we've seen that helps the code run then making it our own.}

In contrast, his CTAS responses to statements such as, ``Generally, I am competent using computational tools (such as spreadsheets, graphing programs, simulations and writing code),'' indicate lower confidence than what we see in his interview. This seems to indicate a difference between his confidence regarding specific tasks (which are well-represented in his COP-Model) versus categories of practices (which are not well-represented).

\subsubsection{Paul: Sustained Motivation with a Nuanced COP-Model}

Paul opened his interview with an interesting contrast, that he sees himself working with computation in the future, even though he fears it comes with ``soul-crushing'' activities:

\quotepar{Paul}{Mostly, I feel like the majority of time [in physics] is just analyzing results on a computer... I'm a computing minor. Um, so I'm getting more and more into code. And I really like programming. So [I will] probably end up in some sort of data analysis. Sounds pretty soul crushing but you do enjoy some of it.}

The interviewer followed up by asking what he meant by ``soul crushing.''
 
\quotepar{Paul}{I just think of an accountant like staring at a spreadsheet, you know, eight hours a day and I don't know, it could get overwhelming. I guess that it depends. Because in physics you're kind of discovering the mysteries of the universe, which isn't very soul crushing. Um, I don't know, I guess, dealing with the mass quantities of data gets a bit much, but I don't know, maybe that comes with time.}

Paul already displays a nuanced COP-Model, where physics requires activities that \textit{could} become soul-crushing, but comes with great reward, and this seems enough to sustain his motivation along an inbound trajectory.

When discussing the benefits offered by computation, Paul identified efficiency and visualization, and again used his COP-Model to sustain his motivation.

\quotepar{Paul}{It's just a way to break the monotony of doing the same equation thousands and thousands of times, you know, a 30-step equation... I'm more of a visual person. So whereas once you'd have a spreadsheet of just data points, now you can actually generate like a histogram or whatnot, and actually know, like, I should be able to... answer questions more easily.}

Paul seemed particularly proud of his computational accomplishments. He described a specific example of when he developed an automated bracketing method for finding eigenergies through repeated use of the shooting method. The additional step of automating this process was not part of the computational assignment, but a result of Paul's COP-Model including efficiency as a reason to use computation. This type of transfer is a goal of computationally integrated instruction \cite{caballero2012implementing}. 

Paul seems to have developed a fair amount of confidence from the MWPs, and expects his confidence to grow further.

\quotepar{Paul}{You're not gonna have a lot of trouble learning Python and the way [the instructors] gave us most of the code and we had to do little things, we didn't really need much programming knowledge... I was on Google a lot just trying to find how to do these things... I got a little more comfortable... I feel like once I got comfortable at [my future] job, if I'm doing something, you know, a thousand times, I'm just going to say, ``Oh, I wish I had a program for this'' and I'd probably just go about making one.}

In contrast with Guy, Paul does not cast googling answers in a negative light, indicating a different representation of the community's standards in his COP-Model.

\subsection{\label{subsec:Themes}Emergent Interview Themes}

Having built a profile of each student's COP-Model, we now examine themes that emerged across student interviews about the use of computation as a physics practice. 
In general, we found that these students' COP-Models included mostly realistic expectations about computation and overall positive outlooks toward computation as a physics practice; however, we also observed shortcomings in their confidence that warrant attention in future courses.

\subsubsection{These students saw computation as part of physics practice.}

When asked Interview Question 1 (``When you picture a physicist conducting research, what kinds of activities do you imagine them doing?''), all five students included the use of computers in their answers. We recognize that the students knew this interview would focus on the use of computation in physics, but we did not previously mention computation in the interview questions. Therefore, this trend seems to indicate that these students' COP-Models of physics included computation at least to some degree after their computationally integrated coursework concluded. In terms of Figure \ref{fig:ComputationPieCharts}, their internal representations of physics practice more closely match the balanced pie chart on the left. In terms of Figure \ref{fig:COP-Model}, computation holds a prominent place in the list of practices that represents the sense of joint enterprise.

\subsubsection{These students saw personally relevant benefits of using computation.}

Collectively, the five students we interviewed identified various important benefits of using computation that overlap with the physics community's reasons for using computation as outlined in Section \ref{sec:IntroComp}. These benefits included visualization, efficiency, accuracy, data analysis, and sense-making. We also saw that these students attributed personal importance to these benefits, as opposed to simply recounting their importance to the global physics community. For example, some students described how the efficiency afforded by computation helps them avoid burnout and focus on physical reasoning over mathematical derivations. 

Looking at the interview excerpts overall, we note that most students prioritized one of these reasons above others in their comments. For example, Chrissie tended to focus on sense-making, Guy tended to focus on data analysis, Harrison tended to focus on efficiency, and Jose tended to focus on visualization. This trend appears to indicate that these students found one benefit of computation to personally resonate with them as they become more comfortable with the practice. Interestingly, Paul mentions efficiency, sense-making, and visualization with roughly equal importance, and his interview answers demonstrate one of the most positive outlooks toward computation, in general. 

We take these features to indicate that these students' COP-Models are beginning to conform to the local COP in terms of the roles that computation plays in the physics community's sense of joint enterprise. It seems likely that Paul could be considered closer to the COP's center than the other students, but they all seem to have a fairly clear and unique inbound trajectory and their COP-Models generally motivate them to persist along their distinct inbound trajectories. Supporting diverse valid trajectories in this way is an important aspect of holistic, learner-centered assessment, which we discuss further in Section \ref{subsec:Recommendations}.

\subsubsection{These students struggle with confidence with regards to computation.}

One need that arose frequently in student interviews is a lack of confidence with regards to their ability to carry out computational tasks and understand computational practices. In particular, four of the five students mentioned a lack of programming experience as either contributing to their low confidence or as posing a challenge for their classmates. It seems that, even though all three instructors attended to the need to orient programming novices at the beginning of the semester, these students still felt that they were on shaky ground. Their COP-Models paint a picture that they feel at least partially unprepared for the next steps in their trajectory.

Students mentioned a deficient background even though most of them reported completing at least one programming course (see Table \ref{table:students}). They highlighted the need to learn a new programming language as a struggle, especially with Python (the language used in their physics courses) functioning so differently than JavaScript or C++ (the languages used in their programming courses). This reported experience runs counter to the advice frequently touted that learning a second programming language is straightforward. We discuss this issue further in Section \ref{subsec:Recommendations}.

After completing physics coursework that relied on the use of MWPs, most of these students felt particularly unprepared to develop new code from scratch. This concern is ironic, given sentiments expressed by the instructors that students should not need to code from scratch, a perspective shared by the global physics community  \cite{thornton2009computational,enrique2018computational,oleynik2019scientific,weatherford12student}. 

We take these differences as examples of how these students' COP-Models do not fully align with the local academic COP (or, by extension, the global professional COP), as they expect to need a skill not held as important by the COP. Still, the question of how the use of MWPs might undercut students' confidence is worth considering further.

\subsubsection{These students wanted computation earlier in the curriculum.}

When asked for feedback about the placement of computation within the physics curriculum, students offered specific ideas based on their experiences. They generally seemed to agree that computation needs to be reinforced throughout the curriculum, and needs to be introduced in either the introductory physics sequence or in a prerequisite programming course. The former approach is favored by many in the physics education community, as  introductory programming courses offered by computer scientists are often oriented differently than the needs experienced in physics courses, because of the differences between these two communities of practice. Indeed, the students' lack of confidence after one or two such courses indicates that the latter approach may not be sufficient, as it leads their early trajectory through peripheral participation that is not legitimate in the physics community.

\section{\label{sec:Discussion}Discussion} 

Having presented key elements from our student and instructor interviews, we next present answers to our research questions, outline implications for computationally integrated physics instruction, reflect on our use of the COP-Model construct, and pose questions for further research into students' experience of computation in physics.


\subsection{\label{subsec:Answers}Answers to Research Questions}

Here, we answer our research questions based on our student and instructor interviews.

\begin{enumerate}
    \item How did the instructors' background with computation inform the local COP created in these courses? The instructors' experience with computation in research heavily influenced their instructional decisions about integrating computation. They also attended to the students' need to use computation in graduate study and industry. We therefore conclude that they designed their courses as a local academic embodiment of the global professional physics community of practice.
    
    \item How does computation feature in each student's COP-Model? Each of these students described computation as a normal part of physics practice that professionals frequently use. These students contrasted learning from computational activities with learning from analytical assignments and laboratory activities, indicating that computation holds a distinct place among physics practices in their COP-Models. This trend highlights an important alignment between their COP-Models and the global professional COP. This outcome is encouraging to see after the negative observations discussed in Sections \ref{sec:IntroComp} and \ref{sec:needs}. We suspect that these students' concurrent enrollment in multiple computationally integrated physics courses helped to reinforce these developments within their COP-Models. 

    \item Where does each student see themselves in the professional physics COP, as assessed by their use of computation? These students' confidence levels grew at varying rates across their different course enrollments depending on the degree to which their legitimate participation was peripheral. They all seemed to end the fall 2020 semester with a sufficient comfort level working with the MWPs from their classes. However, they specifically reported a lack of confidence with developing code from scratch and feeling overly reliant on internet searches for answers. This lack of confidence negatively impacts their sense of their trajectory, and highlights a misalignment between their COP-Model and the professional COP, where such independence is not highly prioritized. Each student described specific opportunities either in a class or in a research experience to use computational practices to their benefit. They also generally see computation as relevant to their future careers, and expect to use it comfortably and frequently. This trend highlights an alignment between their COP-Model and the global COP, and indicates a preparation to transfer computational knowledge to new domains \cite{caballero2012implementing}.

\end{enumerate}

\subsection{\label{subsec:Recommendations}Implications for Instruction}

These interviews have yielded deep insights into these students' experiences. Based on the themes discussed in the previous section, we suggest the following implications for computationally integrated physics instruction:

First, we note the diversity displayed in how the students' COP-Models represent computation as a physics practice. Each student's model prioritizes a different subset of benefits from using computation, which suggests that their success at a computational assessment might depend on the assessment's alignment with their COP-Model. For example, if each of these students were asked to complete a computational project that focused on producing a visualization, Jose and Paul might be more likely to engage with the project, while Guy and Harrison might feel disengaged, preferring to address more technical objectives. On the other hand, an assignment to optimize a code's performance would directly appeal to Harrison's interest but leave the other students uninterested. Such varying levels of engagement might lead to varying levels of performance, which could become coupled to issues of representation. We therefore recommend that computational assessments leave room for students to explore and express their interests, as discussed by Means and Stephens \cite{means2021cultivating}.

Second, we recommend reexamining the frequently presented advice that learning a second programming language is straightforward. The rationale behind this proposition seems innocuous: When a student learns their first programming language, they must learn programming practices as well as the syntax of the language, while learning a second language requires only learning the new syntax. However, this goes against the experience reported by these students, who described in detail challenges they faced when switching from JavaScript or C++ to Python. Attending to this misalignment (now on the part of the central members of the COP) will help improve student confidence.

Finally, we think it wise to examine the ways in which using MWPs might unintentionally undercut students' confidence with programming. Using MWPs certainly helps students meet the real expectations they will face in the professional COP, but we have observed that students' COP-Models can place a misaligned priority on developing code from scratch without internet searches. When instructors observe shortfalls in student confidence with programming, it would be worthwhile to investigate the expectations they are holding themselves to as represented in their COP-Models.

\subsection{Our Use of the COP-Model Construct}

In Section \ref{sec:COPModel}, we outlined the COP-Model as a construct for understanding students' internal representation of a professional community of practice. Here, we reflect on the role this construct played in our study and how these interviews lead us to affirm or revise the COP-Model construct.

\textit{A learner's COP-Model includes the goals and practices that make up their understanding of the COP's sense of joint enterprise.} This aspect of a COP-Model helped us design our interview questions to focus on students experience using physics practices (such as building computational models and debugging) to achieve goals valued by the physics community (such as visualization and extracting insight). Identifying these practices and goals also helped us develop themes while coding our interview transcripts. By considering how practices and goals are represented in a student's mental model, we distinguished between how practices and goals are established by central experts and how they are experienced by novices during legitimate peripheral participation. The trend of students' focusing on one or two goals for using computation prompts us to further consider what emphasis a student's COP-Model places on goals and practices. In the imagery of Figure \ref{fig:COP-Model}, we might say that the student's list representing the sense of joint enterprise has some items written in larger font, closer to the top of the list.

\textit{A learner's COP-Model includes their sense of membership within the COP.} We envisioned the student's perceived position and trajectory as being informed by the degree to which the COP's goals align with their own interests and how the COP's practices align with their own confidence. We found these dimensions of interest and confidence to be useful metrics in forming profiles of our students' COP-Models. We also observed that the students' confidence can be confounded in the COP-Model in two ways: First, student confidence might be based on expectations represented in the COP-Model that are not held in reality (in this case, coding from scratch without internet searches). In such a case, the student's perceived distance from the center might be greater than their distance in reality. Second, student confidence within a community can develop at different rates depending on the degree to which their legitimate participation is peripheral (in this case, related to the degree of scaffolding featured in computational activities). This leads us to wonder how these competing rates of confidence development combine in the space of the student's trajectory $\vec{r}(t)$. In physics terms, we wonder, ``What is the dimensionality of $\vec{r}(t)$, and how do its components interact to form $|\vec{r}(t)|$?'' 

\textit{A learner develops their COP-Model in response to legitimate peripheral participation and feedback.} This aspect of the COP-Model prompted us to order our interview questions chronologically: We started with their fall 2020 computational experiences and progressed to their present opportunities to use computation, and concluded with their expectations of using computation in the future. Doing so allowed us to trace the impact of the students' prior participation to the current outlook offered by their COP-Models. This structure specifically gave us a picture of the trajectory $\vec{r}(t)$ represented in their COP-Model, and the next steps $\Delta \vec{r}$ they expect to take along that trajectory. As noted above, varying degrees of periphery-appropriate participation can complicate these trajectories.

\textit{A learner's COP-Model enables them to extrapolate their experience of a local COP to an understanding of the global COP.} We observed this extrapolation in our student interviews as students related their prior experiences in the local academic COP to the global professional COP. We observed them rely on their COP-Models to envision the activities of professional physicists and describe their own future physics-related careers. We identified points of alignment and misalignment between their COP-Model and the global COP, which we used to build a profile of each student's COP-Model. None of these students described an overall negative COP-Model, so there is additional space to explore with students of a greater diversity of outlooks.

\textit{Comparing models of different COPs helps a learner develop and negotiate their nexus of multimembership.} Our research questions did not focus on nexus of multimembership, so we maintain this point tentatively. The students did compare experiences between their computationally integrated physics courses and their introductory programming courses, which are two different academic communities with different goals and practices. However, these discussions did not deeply probe their use of different COP-Models in these contexts.

Overall, we are satisfied with the insights afforded by this construct, and plan to explore further uses of the COP-Model. We are particularly interested in developing more detailed means of assessing or describing a student's COP-Model. For example, if the COP-Model is a map, can students draw one, or otherwise describe it in a way that can inform their instructors' teaching? Can we assess the degree of alignment between a COP-Model and the COP-in-reality, or at least highlight differences between a novice's COP-Model and an expert's COP-Model? Can we assess the students' perceived distance from the center of the COP?

\subsection{Students' Experience of Computation in Physics}

At the beginning of the paper, we brought up the question, ``How do we know whether computational integration is accomplishing all the promises outlined in Section \ref{sec:IntroComp}?'' Our interviews show evidence of some successful fulfillment of these promises, but the question warrants investigation at a broader scale.

What legitimate peripheral participation do we need to offer students when they first learn programming, or when they learn a new programming language? How are students' COP-Models impacted by the different community entry points of introductory programming courses, computationally integrated introductory courses, or brief introductory tutorials?

Our interviews showed overall positive perceptions of computation, while the anecdata in Sections \ref{sec:IntroComp} and \ref{sec:needs} suggest a much more negative outlook. What are we to make of these differences between students' COP-Models? One important distinction is that much of the anecdata likely comes from the introductory context, while our interviews were confined to the upper-division context. We are also limited by our self-selected pool of student interviewees, who are likely more engaged than others in the population. Once we have a better understanding of these differences between students' COP-Models, we can better address how to motivate computation for our students.

How do students navigate the use of computation across their nexus of multimembership? As computation becomes more broadly integrated across STEM, we will need to attend to differences between how computation is integrated in physics and how it is integrated in other subjects or in our students' future careers. With a better idea of our students' nexus of multimembership, we can better prepare them for using computation in these communities.

How are students' COP-Models shaped by different pedagogical implementations of computation? Most PER work in computation focuses on simulation or modeling, but to what degree do those insights extend to data analysis? Does it make a significant difference whether students obtain the data themselves, or retrieve it from an external source?

\section{Conclusions}

We have explored students' experience in computationally integrated upper-division physics courses using a Communities of Practice framework. Conducting interviews of students and their instructors, we attended to the students' internal representations of how the physics community uses computational practices. We found that these students' internal representations partially align with features of the global professional community, reflecting the diverse trajectories these students are navigating within that community. We note that they struggle with confidence with regards to using computation, partly due to a lack of preparation and partly due to misaligned expectations with the physics community. We have presented recommendations to further support computationally integrated student learning.

\section{Acknowledgements}

We are grateful to our interviewees for their time and insightful conversation. 

\section{\label{sec:protocols}Appendix: Interview Materials}

Items in square brackets were filled in based on the subjects' pre-interview survey and interview responses.


\subsection{Student Pre-Interview Survey\label{subsec:survey}}

Below are a few questions we'd like you to answer before your interview begins. We'll use your answers to customize your interview. 

\textbf{Pseudonym}

During your interview, we'll refer to you by a pseudonym (false name) to help us keep your responses confidential. What pseudonym would you like us to use?

\textbf{Background Questions}
 
In this section, we'll ask you questions about your background using computation in physics.

Please list all the physics classes in which you've used computation in assignments, class activities, projects, or other learning activities. By “computation,” we mean the use of a computer programming environment (Jupyter, python, C++, JavaScript, etc.) to study a technical problem. Please include...

\begin{itemize}
\item The class name (``Modern Physics'') or number (``PHY 3101'').
\item In what semesters (Fall 2020, Spring 2019, etc.) did you take each of these classes?
\item Who were your instructors in each of these classes?
\item On average, approximately how many hours did you spend each week on computational activities during each class?
\end{itemize}

Briefly tell us about any other learning experiences you've had involving computation (other classes, a training session or bootcamp, following on-line tutorials, etc.).

Tell us about your career goals. What type of work would you like to do after you graduate? What type of job would you like to have? What types of skills would you like to use or what types of activities would you like to be involved in?

(The CTAS questions \cite{megowan-romanowicz_2020} appeared here.)

\subsection{Student Interview Protocol}
I'm going to start with some questions about physics research.

1. When you picture a physicist conducting research, what kinds of activities do you imagine them doing? 
How do you think [activities] relate to each other?
Which of [activities] do you see yourself doing in the future?
Ask other probing questions as appropriate. 

2. Why do you think physicists use computation in their research?
How do you think computation relates to [activities]?
Ask other probing questions as appropriate.

Next I'm going to ask some questions about your experience with computation in physics.

3. In [classes], what was one important physics concept you learned from your computational assignments? [Record their answer below.] 
[concept] = 
How did the computational assignments help you learn [concept]? 
Was [concept] something you already knew before the computational assignment, or was it something the computational assignment showed you for the first time?
Ask other probing questions as appropriate.

4. IF Modern Physics in [classes]: Specifically, in Modern Physics, how did the computational assignments help you learn about the weird concepts in quantum mechanics? 
[Give examples of “weird concepts” as needed: particle-wave duality, tunneling, uncertainty principle]
Can you describe an example? 
How did the computational assignments help differently than other aspects of the class, like reading the textbook or solving problems?
Ask other probing questions as appropriate.

5. What sort of computational skills, if any, do you think every physics student should learn about?
Ask probing questions as appropriate.

6. Describe how comfortable you felt working through the computational assignments in [classes]. 
Ask probing questions as appropriate.
Questions about Current thoughts about Computation in Physics
Next, I'm going to ask some questions about your current thoughts about computation in physics.

7. Suppose you had to go back today and work a little more on the computational assignments from [classes]. Describe how comfortable you would feel now working through those computational assignments. 
Why do think you'd feel that way?
Ask probing questions as appropriate.

8. Is there a class you're taking now where you might use computation or learn more about computation? 
Tell me about that.
How comfortable would you/do you feel using computation or learning more about computation in that class?
Ask other probing questions as appropriate.

9. Are you working on any research projects right now where you might use computation? 
[If they say no or seem reluctant, you might want to broaden the definition of “research” to be “any study or investigation where you're trying to learn something new, either in class or out of class.”]
Tell me about that.
How comfortable would you/do you feel using computation in that project?
Ask other probing questions as appropriate.

10. When you see your future self as a [career goal], how do you see yourself using computation, if at all?
IF affirmative answer: How comfortable do you think would you feel using computation like that? 
IF negative answer: Why is that?
Ask other probing questions as appropriate. Specifically probe about…
How frequently they expect to use computation (regularly, infrequently, rarely).
To what degree they expect to use computation (central to their work, as needed to support their work).
In what ways they expect to use computation (modeling, data analysis, visualization).

11. What feedback or suggestions would you give to [instructors] about the computational assignments in [classes]?
What aspects of the assignments worked well?
What aspects of the assignments didn't work for you?
What frustrated you about the assignments?
How would you improve the experience?
Ask other probing questions as appropriate.

I have one final question.

12. Is there anything else you'd like to go back to or add to our conversation?
Ask probing questions as appropriate.

\subsection{Instructor Interview Protocol}
1. What field of physics do you conduct research in? When you conduct research, what methods do you use?  How would you describe the usefulness of each of these methods for your research? How useful would you rank computation?

2. How do you use computation in your research, teaching, or any other professional work? How often do you use computational physics methods for [see list above]? In what courses do you use computational physics methods? = [courses] What computational platform(s) do you use? What results does this platform yield? [If necessary, give examples: Jupyter notebook, Excel, Logger Pro, etc.] How do you integrate computation into your classes? Can you describe an example of a computational assignment? 

3. For what learning objectives, if any, do you use computational activities to support? Why? What computational platforms do you use? Why did you choose this platform? What makes computational physics methods more appealing for this concept rather than analytical or experimental activities? How do you use computational physics methods for this concept? How do computational physics methods allow you to make and analyze models? Why would this be important? How are computational physics methods useful for concepts that are hard to visualize (i.e. imaginary numbers, normal force, sound waves, anything that you can't visibly see)? 

4. What sort of computational skills, if any, do you think every physicist should have? = [skills] Why is [skill] important? Where is [skill] used?  When do you think these skills should be taught? Why? How well do you think your students develop these skills?

5. Describe how successful your use of computation has gone in [courses]. What did you find difficult? Why? What did your students find difficult?  What did you find easy, if anything? Why? What did your students find easy? Overall, how would you describe your students' learning from their computational assignments? What computational skills did you want your students to take away from your [course] class? 


\bibliography{apssamp}

\end{document}